\documentclass[twocolumn,showpacs,preprintnumbers,amsmath,amssymb]{revtex4}

\usepackage{slashed}
\usepackage{graphicx,epsfig}
\usepackage{dcolumn}
\usepackage{bm}

\newcommand{\ga}{\alpha}
\newcommand{\gb}{\beta}

\newcommand{\go}{\omega}

\newcommand{\gL}{\Lambda}

\begin{document}

\preprint{}

\title{Light front QED$_{1+1}$ at finite temperature}

\author{S.~Strauss}
    \email{stefan.strauss@uni-rostock.de}
\author{M.~Beyer}%
 \email{michael.beyer@uni-rostock.de}
\affiliation{%
Institute of Physics, University of Rostock, 18051 Rostock, Germany\\
}%

\date{\today}

\begin{abstract}
We investigate thermodynamic properties of quantum
electrodynamics in $1+1$ dimensions (QED$_{1+1}$)
utilizing light front dynamics. Therefore we
derive the partition function of the canonical ensemble
in discrete light cone quantization, and
calculate the thermodynamical potential.
This central quantity is evaluated for
different system sizes and coupling strengths.  We investigate the
continuum limit and the thermodynamical limit and
present basic thermodynamical quantities
such as pressure, energy, and entropy, as a function of temperature for the
interacting system. The results are compared to the ideal bosonic
and fermionic cases.

\end{abstract}

\maketitle

Recently, Light Front Quantization (LFQ) originally introduced by
Dirac in 1949~\cite{Dirac:1949cp} has been successfully extended to
relativistic systems of finite temperatures $T$, and also chemical
potentials
$\mu$~\cite{Elser:1996tq,Beyer:2001bc,Alves:2002tx,Weldon:2003uz,Kvinikhidze:2003wc,Dalley:2004ca,Raufeisen:2004dg}.
The interest in utilizing LFQ in the area of quantum statistics and
thermo-field theory is motivated through the need to investigate
strongly coupled relativistic systems.  A pertinent example of such a system is
given by the quark gluon plasma (QGP), that is a new state of matter
recently discovered at the relativistic heavy ion collider
(RHIC)~\cite{Rischke:2005ne}.

Practical use of LFQ as a nonperturbative method to treat strongly
interacting systems has been realized through Discrete Light Cone
Quantization (DLCQ) introduced in~\cite{Pauli:1985pv}. Subsequently
DLCQ has been used to treat two-dimensional gauge theories
like Quantum Electrodynamics QED$_{1+1}$~\cite{Eller:1986nt,Yung:1991ua} and
Quantum Chromodynamics QCD$_{1+1}$~\cite{Hornbostel:1988fb}. Following these
promising applications of DLCQ many other models of various dimensions
have been investigated.  Further progress in the Light front Schwinger model
has been obtained utilizing the Tamm-Dancoff
approximation~\cite{Mo:1992sv,Harada:1997kq}. Besides the LFQ approach finite
lattice size computations have been performed, see e.g.
Refs.~\cite{Sriganesh:1999ws,Byrnes:2002nv} and refs. therein.
Alternatively, an instant form Hamiltonian lattice approach in a fast
moving frame was shown to give reasonable
results~\cite{Kroger:1998se}. The interest in QED$_{1+1}$ is motivated
by the appearance of phenomena like chiral symmetry breaking, charge
confinement, and topological properties associated to $\theta$ vacua
similar to full QCD. In addition, it is a widely
studied theory to investigate new solution methods like the one
presented here.

In the initial work of Ref.~\cite{Elser:1996tq} that
focused on finite temperatures,
 DLCQ was merely introduced to compute the
mass spectrum of QED$_{1+1}$. Subsequently, the mass
spectrum has then been used in an
instant form framework to determine thermodynamic quantities like the
partition function, internal energy, specific heat, etc. In this
framework the authors obtained remarkable results including a second
order phase transition at around $T\simeq 1\ g/\sqrt{\pi}$
(where the strength $g$ will be specified below). Extrapolation
to the continuum limit  leads to a lower
bound for the critical exponent $\ga > 0.7$ and critical
temperatures $T_c = 0.5\dots1.0\ g/\sqrt{\pi}$ (depending on the interaction)
even in the strong coupling regime, where the Schwinger model is exactly solvable by
bosonization. This is astonishing, since in a correct treatment of the
two limits, (i) weak coupling $m/g\rightarrow \infty$ and (ii) strong
coupling $m/g\rightarrow0$ (where $m$ is the mass unit, see below), a free
fermion respectively a free boson equation of state (EOS) should be
realized.

Within the same framework of~\cite{Elser:1996tq} Hiller et
al. investigated a super-symmetric model in two
dimensions~\cite{Hiller:2004vy,Hiller:2007sc}. They computed
thermodynamic observables of the bosonic and fermionic sectors of the
theory. It turns out that irrespective of the specific findings of
Ref.~\cite{Elser:1996tq} the models under
concern~\cite{Hiller:2004vy,Hiller:2007sc} due to the large $N_c$
limit lead to a free theory of mesonic states. As a consequence the
EOS is trivially that of an ideal system of bosons.

In order to study the non-ideal case we investigate the massive chiral
Schwinger model for interaction terms of finite values of $m/g$.  We
first solve the DLCQ Hamiltonian up to some harmonic resolution and
then evaluate the proper partition function for finite temperature
without introducing any quasi particle concept. The
continuum and the thermodynamic limits are carefully performed.
\begin{figure}
\begin{center}
\epsfig{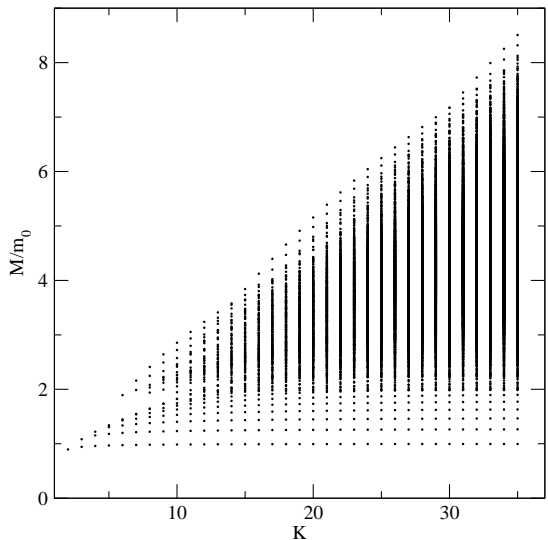}
\end{center}
\caption{\label{fig:M2} Invariant mass spectrum of QED$_{1+1}$ for
  given harmonic resolutions $K$ and $m/g=1$. $m_0$ denotes the lowest
  eigenvalue for $K\rightarrow\infty$ and is used for normalization.}
\end{figure}

We introduce the massive chiral Schwinger model in DLCQ following
Refs.~\cite{Eller:1986nt,Yung:1991ua}. Within the DLCQ approach the
light-like physical extension $x^-$ is restricted to $-L/2 \le x^- \le
L/2$. Therefore by demanding (anti)-periodic boundary conditions of the
fields the conjugate momentum variable $p^+$ becomes discrete $p^+_n=
\frac{2\pi}{L}n,\; n\in\mathbb{N}$. For periodic boundary conditions
the $n=0$ mode (fermionic zero mode) exists while for anti-periodic
boundary condition $n>0$ holds. In the following we choose light cone gauge $A^+=0$
and neglect the fermionic and gauge zero modes to not obscure the thermodynamic
issues addressed here.

The light cone momentum operators can then be
written as
\begin{equation}
\label{eq:p^+,-}
P^+ = \frac{2\pi}{L}K,\qquad P^- = \frac{L}{2\pi}H
\end{equation}
with the harmonic resolution $K$ being dimensionless and the LF
Hamiltonian $H$ of dimension mass squared.
Expressed in Fock space
creation/annihilation operators for particles ($b_n^\dag$) and
antiparticles ($d_n^\dag$) the harmonic resolution and the free
(dimensionless) Hamiltonian $H_0$ are
\begin{equation}
\label{eq:resolution}
K = \sum\limits_{n=1}^\gL n\left(b_n^\dag b_n + d_n^\dag d_n\right),
\end{equation}
\begin{equation}
\label{eq:p^+}
H_0 = \sum\limits_{n=1}^\gL \frac{1}{n}\left(b_n^\dag b_n + d_n^\dag d_n\right)
\end{equation}
with a high momentum cutoff $\gL$. The full Hamiltonian is given
by~\cite{Eller:1986nt}
\begin{equation}
\label{eq:hamilton}
H = m^2 H_0 + \frac{g^2}{\pi}V = g^2(\frac{m^2}{g^2} H_0 + \frac{1}{\pi}V),
\end{equation}
where $m$ is the bare mass and $g$ the interaction strength. We have
introduced the ratio $m/g$ that for $m/g\rightarrow 0$ gives the
strong and $m/g\rightarrow \infty$ the weak coupling limit. The Fock
space representation of the potential $V$ is a rather lengthy
expression that has been given in Ref.~\cite{Eller:1986nt} Sec. 2, and
will not be repeated here for the sake of brevity.

\begin{table}[b]
\caption{\label{tab:numericalcomp} The different estimates for the
  bound state energies $E_1/g$, $E_2/g$ of the lowest two mass
  eigenstates at various $m/g$.  For comparison the results of
  \cite{Mo:1992sv} and \cite{Eller:1986nt} using LCQ, finite-lattice
  results of \cite{Byrnes:2002nv} (first state),
  \cite{Sriganesh:1999ws} (second state), and data obtained by the
  fast-moving frame lattice Hamiltonian method~\cite{Kroger:1998se}
  are shown. Error given due to variation of extrapolating functions
(The case $g/m=2^0$ is also shown in Fig.~\ref{fig:mass}).}
\begin{ruledtabular}
\begin{tabular}{cccccc}
 $m/g$& this work &~\cite{Mo:1992sv} &~\cite{Eller:1986nt} & ~\cite{Sriganesh:1999ws},~\cite{Byrnes:2002nv}&~\cite{Kroger:1998se} \\
 \hline
 \multicolumn{5}{c}{ground state / vector state} \\
 $2^5$&0.191(3)&0.201&0.201&0.194(5)&0.191\\
 $2^4$&0.2366(8)&0.224&0.228&0.238(5)&0.235\\
 $2^3$&0.2856(4)&0.288&0.280&0.287(8)&0.285\\
 $2^2$&0.33933(5)&0.337&0.338&0.340(1)&0.339\\
 $2^1$&0.39355(4)&0.393&0.393&0.398(1)&0.394\\
 $2^0$&0.4442(7)&0.444&0.444&0.4444(1)&0.445\\
 $2^{-1}$&0.4873(1)&0.488&0.488&0.48747(2)&0.489\\
 $2^{-2}$&0.519(1)&0.520&0.534&0.51918(5)&0.511\\
 $2^{-3}$&0.538&0.540&0.603&0.53950(7)&0.528\\
 \hline
 \multicolumn{5}{c}{first excited state  / scalar state} \\
 $2^5$&0.46(7)&0.458&0.458&0.45(1)&0.447\\
 $2^4$&0.5623(3)&0.564&0.548&0.56(1)&0.559\\
 $2^3$&0.696(4)&0.697&0.689&0.68(1)&0.690\\
 $2^2$&0.839(2)&0.838&0.839&0.85(2)&0.837\\
 $2^1$&0.9892(1)&0.989&0.985&1.00(2)&0.991\\
 $2^0$&1.117(1)&1.119&1.126&1.12(3)&1.128\\
 $2^{-1}$&1.2002(2)&1.201&1.228&1.20(3)&1.227\\
 $2^{-2}$&1.21(3)&1.230&1.312&1.24(3)&1.279\\
 $2^{-3}$&1.27(1)&1.219&1.407&1.22(2)&1.314\\
\end{tabular}
\end{ruledtabular}
\end{table}

We first determine the invariant mass spectrum $M^2=P^+P^-=KH(K)$. To do so,
we construct the Fock space for a given resolution $K$ and collect all
possible integer partitions $\{K\}$, where integers represent a
single-particle momentum state of an electron or positron for a given
momentum while the Pauli principle is observed. The Hamiltonian is
block-diagonal in $K$ and the resulting Fock space spanned by $\{K\}$
is finite.  We then calculate the DLCQ QED$_{1+1}$ Hamiltonian matrix
$H(K)$, which can be diagonalized for values of $K\lesssim 35$ without
much effort on today standard PCs. The mass spectrum $M=\sqrt{KH(K)}$
is shown in Figure~\ref{fig:M2} as a function of $K$.

\begin{figure}
\begin{center}
\epsfig{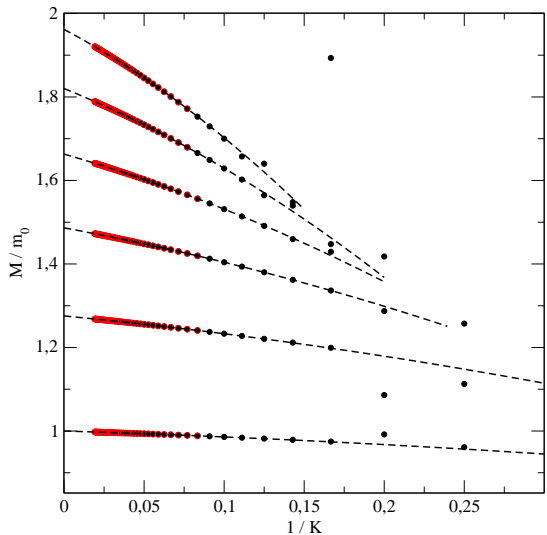}
\end{center}
\caption{\label{fig:mass} The six lowest mass eigen values of
  QED$_{1+1}$ as a function of $1/K$ for $m/g=1$. The dashed line is a quadratic
  fit to the data (values used in the fit are colored in red) and to extract the
  continuum limit.}
\end{figure}

At larger $K>35$ the dimension of the mass matrix soon becomes rather
large. We therefore restrict ourselves to the lower part of the mass
spectrum.  To this end we introduce standard Krylov space methods
to determine the lowest
eigenvalues as used e.g. in Fig.~\ref{fig:mass}.

Since $M^2=KH(K)$ does not explicitly depend on $L$, the continuum
limit is achieved by $K\rightarrow\infty$ which in turn implies for any
given but (due to covariance) arbitrary $P^+>0$ that
$L\rightarrow\infty$ (continuum). Assuming some low order
fits explained below the continuum
mass spectrum including an educated estimate of the error related to
variation of the fit function or/and parameters is read off at
$1/K\rightarrow 0$. This is demonstrated in Figure~\ref{fig:mass} for
the binding energy $E=M-2m$ of the six lowest mass eigenvalues $M$ for
the case $m/g=1$. In Table~\ref{tab:numericalcomp} we compare our
results with earlier
calculations~\cite{Mo:1992sv,Eller:1986nt,Sriganesh:1999ws,Byrnes:2002nv,Kroger:1998se}
that have been given for the lowest two mass eigenvalues, the ground
(vector) state and the first excited (scalar) state, for
different interaction strengths $m/g$.

We use different extrapolation functions to extract the low lying
masses depending on the coupling regime $m/g$ to arrive at our results
of Table~\ref{tab:numericalcomp}:  At low coupling (that
is $m/g \ge 2^5$) the masses vary strongly and follow a linear curve
in $1/K$, which is used in the fit, only for rather
large resolutions $K\ge 40$.  In the
intermediate coupling regime $2^{-1}\ge m/g\ge 2^4$ the light cone
approach produces the most accurate results even for moderate $K$.
However, a linear continuum extrapolation
systematically gives too high estimates. Thus omitting the linear fit
we do the analysis using second order polynomials. At small $m/g$ ratios
the scaling behavior is considerably different from the previous ones.  Even at
large resolution $K$ the computed masses are still quite far away from
the very precise continuum values found in~\cite{Byrnes:2002nv,Sriganesh:1999ws}.
Changing the fit function
to a polynomial in $1/\sqrt{K}$ we obtain reasonable results at
$m/g=2^{-2}$ but unstable ones at $m/g=2^{-3}$. Therefore we do not indicate
errors for the smallest $m/g$ values but simply present our best fit
estimates. All other errors have been determined through successive
fits with next and next-to-next order power function applied to the
same sample of data points. It is a well-known fact that ordinary LFQ
does not capture all physics in the massless limit, e.g. ignores left moving
fermions, that indicates why this approach has difficulties
reaching the continuum.  This can be cured by introducing a second
light front~\cite{Mccartor:1988bc} or near light cone
coordinates~\cite{Lenz:1991sa}.  Furthermore a proper inclusion of the
gauge field zero mode following~\cite{Martinovic:1997va} should improve
the convergence at small fermion mass. 
Interestingly, the bosonization of the massive chiral LC Schwinger model as
suggested in Ref.~\cite{Eller:1986nt} leads to different induced inertias,
but otherwise leaves the interaction $V$ in \eqref{eq:hamilton}
unchanged. The bosonized model gives exact results in the massless
limit, however, even for strong coupling, similar uncertainties are present
in the fermionic approach.

After solving the isolated DLCQ case to the present state of the art precision,
we now turn to a canonical ensemble subject to the DLCQ Hamiltonian of
QED$_{1+1}$ given above. To calculate thermodynamic properties we
evaluate the respective partition function that is the central
quantity. Following the standard quantum statistical approach the
partition function is evaluated by performing a trace over the
statistical operator in the discretized Fock space. In LFQ
the respective trace is given by
\begin{equation}
\begin{split}
\label{eq:partition}
{\cal Z} &= {\rm Tr}\; \exp\{-\frac{\gb}{2} (P^++P^-)\} \\
&= \sum\limits_K
\exp\left\{-\frac{\gb}{2}\frac{2\pi}{L}K\right\}\ \zeta_K,
\qquad \mathrm{where}\\
\zeta_K&=
\sum\limits_{\{K\}}\exp\left\{-\frac{\gb}{2}
\frac{L}{2\pi}\frac{M^2(\{K\})}{K}\right\}.
\end{split}
\end{equation}
The last two lines refer to the DLCQ expression, wherein the first
part is the sum over the complete Fock space, and $\zeta_K$
contains all Fock state partitions belonging to the given resolution
$K$. The inverse temperature $\gb=1/(k_BT)$ ($k_B$ Boltzmann constant)
is the Lorentz invariant rest-frame temperature. In order to keep the
standard definition of the temperature the (however kinematic) $P^+$
component in the proper definition of the statistical operator on the
light front appears in addition of the light front Hamiltonian
$P^-$. Note that only for the interaction free case $(P^++P^-)/2=P^0$
can be replaced by $P^0=\sqrt{\vec{P}^2+M^2}$, where the Lorentz
invariant mass eigenvalues $M$ may be evaluated is any of the
relativistic forms, e.g. light front form, as has been utilized in
Refs.~\cite{Elser:1996tq,Hiller:2004vy,Hiller:2007sc}. The case
considered here, however, includes interactions and hence the
evaluation of the trace becomes more cumbersome.

To give an instructive example, which is also used to estimate the
theoretical error of the calculation, we evaluate the thermodynamic
potential density $\omega_f=\Omega_f/L$ for the canonical
ensemble  of the ideal fermi gas.
The thermodynamical potential for two particle species is given as
\begin{equation}
\label{eq:lc_freepot}
 \go_f = -2T\int\limits_0^\infty
\frac{dp^+}{2\pi}\ln\left(1+\exp\left\{-\gb\left(\frac{p^+}{2}+
\frac{m^2}{2p^+}\right)\right\}\right),
\end{equation}
where we set the mass to $m=0.5 \mathrm{MeV}$.
To compare (\ref{eq:lc_freepot}) to the standard instant form expression,
the r.h.s.  can be directly transformed by a simple variable
substitution.  In the thermodynamic limit, the instant and light front
form give equal results for the ideal case.
\begin{figure}[t]
\begin{center}
\epsfig{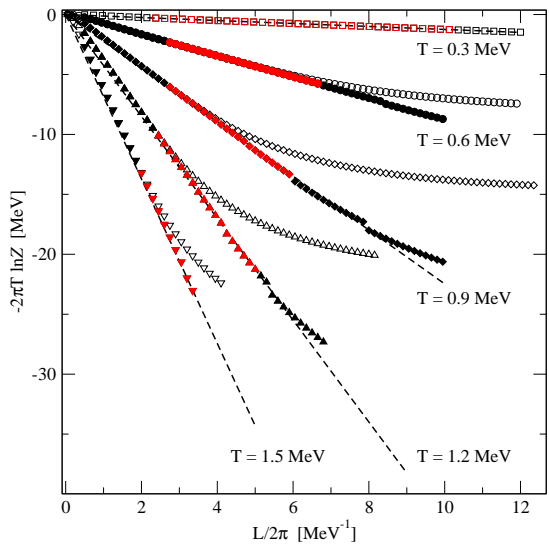}
\end{center}
\caption{The thermodynamical potential $\Omega_f=-2\pi T\ln\mathcal{Z}$
  as a function of $L$. Open symbols depict the potential at maximal
  resolution $K=110$ for the different temperatures. Closed symbols
  are given by the extrapolation as explained in the text. The slope
  of the linear part (values selected colored in red) is fitted to
  extract the invariant potential density.}
\label{fig:freethpot}
\end{figure}
\begin{table}[b]
\caption{\label{tab:lnZ} Values of $\ln{Z}$ for the free fermi gas
 extracted from
  a linear fit to the numerical analysis along with the analytical
  results and the estimated uncertainties.}
\begin{ruledtabular}
\begin{tabular}{cccc}
&\multicolumn{3}{c}{$-2\pi T\ln\mathcal{Z} \mathrm{[MeV]}$} \\
T [MeV]& analytic& numeric&rel.error  [\%]\\
 \hline
0.3 &-0.124156& -0.124263 & 0.05 \\
0.6&-0.862941 & -0.863176&0.03\\
0.9&-2.24692 &-2.24824 & 0.05\\
1.2&-4.24937 &-4.24566 &0.07\\
1.5&-6.85920&-6.76379&1.39\\
 \end{tabular}
\end{ruledtabular}
\end{table}

Evaluating the free thermodynamic potential density
(\ref{eq:lc_freepot}) for a given temperature will be referred to as
the analytical result. To demonstrate our method this
will be compared in the following to the direct
numerical evaluation of (\ref{eq:partition}) for $V=0$ that depends on
the length scale $L$ and a maximum $K$.
Since $\ln\mathcal{Z}$ is an extensive property it scales
with $L$ as well as $\omega=-T\ln\mathcal{Z}$. For a given maximal
harmonic resolution $K$ the numerical results leads to
$\ln\mathcal{Z}$ as a function of the system size $L$.  Hence
$\ln\mathcal{Z}$ can be read off the slope with respect to $L$. The
situation for different temperatures is depicted in
Figure~\ref{fig:freethpot}. The analytic (exact) results are given as dashed
lines and the numerical values as symbols, empty ones reflecting raw
numerical data up to $K=110$. The filled
symbols are raw data corrected by a nonlinear fit as explained below.
For large system sizes the
harmonic resolution used in the evaluation is not sufficient, hence a
deviation from the expected linear dependence of $\ln\mathcal{Z}$ on
$L$ appears. For small $L$ the system is too small to read off the
thermodynamic limit, which also leads to a derivation (finite size effects).
There is an
optimal region of system sizes where $\ln\mathcal{Z}$ depends almost
linear on $L$, which is used to extract the slope by making a linear
fit (indicated by red symbols in Fig.~\ref{fig:freethpot}).
In order to improve the numerical
results we have used the following fitting algorithm for $\mathcal{Z}$ of
(\ref{eq:partition}): Beyond the maximum of $\mathcal{Z}$ we approximate
$\zeta_K$ by an exponential of $\sum_{n=-2}^2a_nK^n$.
The function depends on five parameters $a_n$ that are determined by
a $\chi^2$-fit to the largest 30 calculated $K$-values and
then used for an extrapolation to larger $K$.
The resulting values of $\ln\mathcal{Z}$ for the four temperatures  are given in
Table~\ref{tab:lnZ}. The error indicated is the difference of the numerical
from the analytic result. For small $L$ the analytic
value is approached from above and for large values it deviates again
to larger values. As a consequence the slope (i.e. $\ln\mathcal{Z}$)
is usually slightly too steep.
For the interacting case, where no exact analytic result is available,
we utilize the same fitting procedure.  Note however two
differences to the pure fermionic case. First, particle number is not conserved because of an
interacting field theory and, second, the charge $Q$ of each state of
the statistical ensemble is set to zero due to confinement.  To be
more explicit, the ideal fermionic (bosonic) gas as a limit of
$g/m\rightarrow 0$ ($m/g\rightarrow0$) has been evaluated for
chemical potential $\mu=0$. In the fermionic case this has to be done using the full
numerical calculation obeying the condition $Q=0$.  However, the
bosonic case
can be worked out directly by evaluating the following integral
\begin{equation}
\label{eq:lc_boson}
 \go_b = 2T\int\limits_0^\infty
\frac{dp^+}{2\pi}\ln\left(1-\exp\left\{-\gb\left(\frac{p^+}{2}+
\frac{1}{2p^+}\right)\right\}\right),
\end{equation}
since Bose-Einstein condensation is absent in one
dimension~\cite{Hohenberg:67ll} (and would occur only at $\mu=m$ for an
ideal, relativistic bosonic gas in three dimensions).
\begin{figure}[t]
\begin{center}
\epsfig{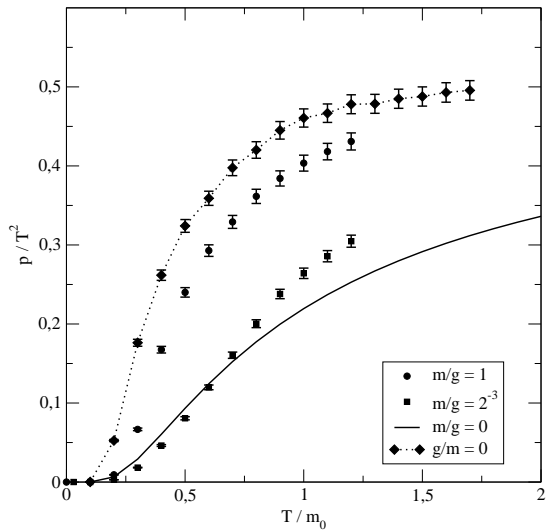}
\end{center}
\caption{Pressure as a function of temperature divided by $T^2$ for
  different models: pure
bose gas (solid line), stronger interacting Fermi system $m/g=2^{-3}$
(squares), weaker interacting Fermi system $m/g=1$ (circles),
noninteracting Fermi gas in $Q=0$ sector as explained in the text
(diamonds). Estimated error as explained in the text.}
\label{fig:druck}
\end{figure}
\begin{figure}[t]
\begin{center}
\epsfig{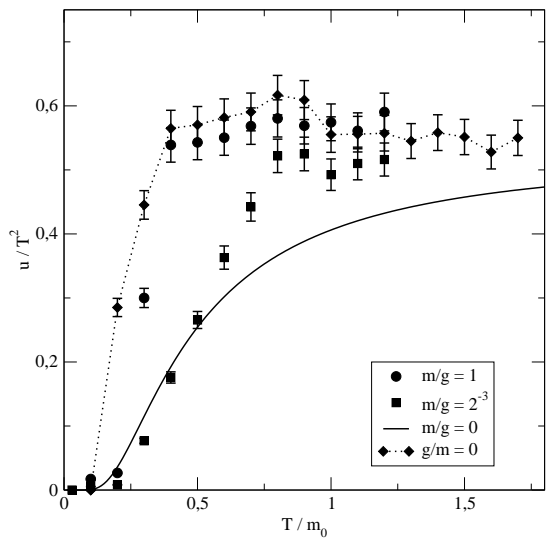}
\end{center}
\caption{Internal energy density of interacting gas as a function of temperature
divided by $T^2$,
for explanation of curves see caption of Fig.~\ref{fig:druck}.}
\label{fig:energy}
\end{figure}
\begin{figure}[t]
\begin{center}
\epsfig{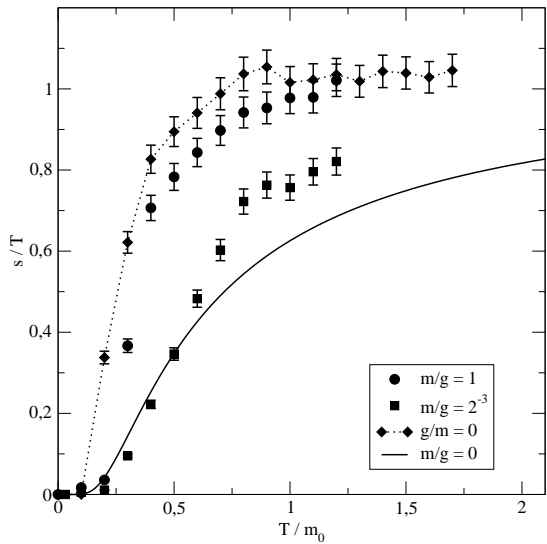}
\end{center}
\caption{Entropy density of interacting gas as a function of temperature
divided by $T$,
for explanation of curves see caption of Fig.~\ref{fig:druck}.}
\label{fig:entropy}
\end{figure}
\begin{figure}[t]
\begin{center}
\epsfig{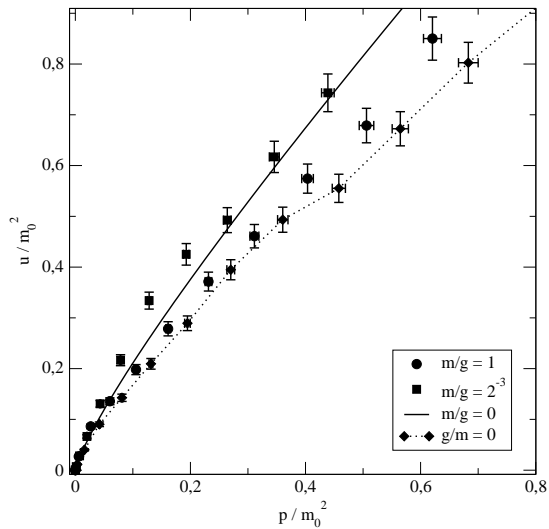}
\end{center}
\caption{\label{fig:eos} Equation of state, pressure $p$ vs. energy density
$u$ for QED$_{1+1}$,
for explanation of curves see caption of Fig.~\ref{fig:druck}.
 The data   points belong to the following temperatures $T/m_0=0.3,0.6,0.9,1.2,1.5$.}
\end{figure}

In the reminder we give several thermodynamic quantities. Note
that volume and temperature are now given
in units of the ground state mass $m_0$ that sets the scale
of the interacting system. In
Fig.~\ref{fig:druck} we show the pressure
$p=-\omega=T\ln\mathcal{Z}/L$ plotted as a function of
temperature. Division is by $T^2$ (rather than the usual $T^4$) is due
to the reduced dimension of the problem.  The line shows the analytic
result for the ideal bosonic case (limit $m/g\rightarrow 0$) according
to (\ref{eq:lc_boson}), the diamonds the ideal fermionic case (limit
$m/g \rightarrow \infty$). The other
symbols represent two different interacting cases. The strong
interacting one $m/g=2^{-3}$ (squares) is closer to the bosonic case,
whereas the weak interacting one $m/g=1$ (circles) is closer to the ideal
fermionic case.  There is a smooth transition from the bosonic case
to the fermionic case, depending on the interaction strength. In particular,
we do not observe a critical behavior seen in \cite{Elser:1996tq} that might have
been due to the rather low harmonic resolution used there.
For $T\rightarrow \infty$, i.e. the classical ultrarelativistic case,
masses can be neglected $m\rightarrow 0$ and hence all curves approach
$p/T^2=\pi/6$. From Table~\ref{tab:numericalcomp} we
assume a conservative mean error of 2.5\% for the
interacting case indicated in figures by error bars.

The energy density $u$ is achieved by thermodynamic relations
involving a numerical derivative of $\ln\mathcal{Z}$, hence error bars
indicated are bigger.  With the same coding $u/T^2$ is shown in
Fig.~\ref{fig:energy}. In the classic ultrarelativistic limit the energy density converges
to $u/T^2\rightarrow \pi/6$.

The entropy density $s=(p+u)/T$ divided by $T$ is shown in
Fig.~\ref{fig:entropy} with the classical ultrarelativistic limit $s/T\rightarrow \pi/3$.

The resulting equation of state is shown in
Fig.~\ref{fig:eos}. For moderate temperatures $p\simeq 2 u$, which is taken
to estimate the error in $u$ to be
twice that of $p$.

In conclusion, we have presented for the first time a possible way to
directly evaluate the partition function of a strongly interacting
relativistic system within the framework of light front
quantization. To this end we have utilized DLCQ that leads to a
Hamiltonian evaluated in a Fock space of finite large dimension. As an
interaction we have chosen QED$_{1+1}$. For illustration we have
investigated some intermediate cases between the noninteracting
(fermionic) limit (chemical potential $\mu=0$, charge $Q=0$
sector) and the completely coupled (bosonic) limit of QED$_{1+1}$.

Present numerical effort is manageable, i.e. all calculations have
been performed on a regular PC. It is also desirable that
besides evidence of feasibility further thermodynamic quantities
could be calculated, such as specific heat or speed of sound. Some
additional effort is needed here, since either for use of thermodynamic
relations (numerical)  derivatives are involved, or some quantities
like energy $U=\langle H\rangle$  have to be calculated from
thermodynamic averaging. In any case within this framework
it is also possible to evaluate correlation functions and determine thermal
masses in a next step.

In view of demands to investigate the QCD phase transitions,
the method can now be extended to QCD$_{1+1}$, e.g., along the lines
of~\cite{Hornbostel:1988fb}. A further next step along this line would be the
inclusion of transverse degrees of freedom, and finally the extension
to finite chemical potentials which is by construction not biased by
fermion doubling and sign problems due to the generic Hamiltonian
approach.

We gratefully acknowledge discussions with T. Frederico during mutual
research stays at each others institutions. Work supported by the
Deutsche Forschungsgemeinschaft (DFG) under contract BE1092/13.

\bibliography{qedfT}

\end{document}